\documentclass[prd, twocolumn, lengthcheck, superscriptaddress, showpacs, letterpaper, nofootinbib]{revtex4}

\usepackage{ifpdf}
\usepackage{color}
\usepackage{graphicx}  
\usepackage{float}
\usepackage{amsmath}
\usepackage{acronym}
\usepackage{multirow}
\usepackage{xspace}
\usepackage{units}
\usepackage{aas_macros}
\usepackage{hyperref}
\usepackage{ulem}

\begin{document}

\title{Reconstructing the sky location of gravitational-wave detected compact binary systems: methodology for testing and comparison}

\author{T. Sidery}
\email{tsidery@star.sr.bham.ac.uk}
\affiliation{School of Physics and Astronomy, University of Birmingham, Birmingham, B15 2TT, UK}
\author{B. Aylott}
\affiliation{School of Physics and Astronomy, University of Birmingham, Birmingham, B15 2TT, UK}
\author{N. Christensen}
\affiliation{Physics and Astronomy, Carleton College, Northfield MN 55057, USA}
\author{B. Farr}
\affiliation{Department of Physics and Astronomy \& Center for Interdisciplinary Exploration and Research in Astrophysics (CIERA), Northwestern University, Evanston, IL 60208, USA}
\affiliation{School of Physics and Astronomy, University of Birmingham, Birmingham, B15 2TT, UK}
\author{W. Farr}
\affiliation{Department of Physics and Astronomy \& Center for Interdisciplinary Exploration and Research in Astrophysics (CIERA), Northwestern University, Evanston, IL 60208, USA}
\affiliation{School of Physics and Astronomy, University of Birmingham, Birmingham, B15 2TT, UK}
\author{F. Feroz}
\affiliation{Cavendish Laboratory, University of Cambridge, Cambridge, UK}
\author{J. Gair}
\affiliation{Institute of Astronomy, University of Cambridge, Cambridge, UK}
\author{K. Grover}
\affiliation{School of Physics and Astronomy, University of Birmingham, Birmingham, B15 2TT, UK}
\author{P. Graff}
\affiliation{NASA Goddard Space Flight Center, Greenbelt, MD, USA}
\author{C. Hanna}
\affiliation{Perimeter Institute for Theoretical Physics, Waterloo, Ontario N2L 2Y5, Canada}
\author{V. Kalogera}
\affiliation{Department of Physics and Astronomy \& Center for Interdisciplinary Exploration and Research in Astrophysics (CIERA), Northwestern University, Evanston, IL 60208, USA}
\author{I. Mandel}
\affiliation{School of Physics and Astronomy, University of Birmingham, Birmingham, B15 2TT, UK}
\author{R. O'Shaughnessy}
\affiliation{Center for Gravitation and Cosmology, University of Wisconsin-Milwaukee, Milwaukee, WI 53211, USA}
\author{M. Pitkin}
\affiliation{SUPA, School of Physics and Astronomy, University of Glasgow, University Avenue Glasgow, G12 8QQ, UK}
\author{L. Price}
\affiliation{LIGO - California Institute of Technology, Pasadena, CA 91125, USA}
\author{V. Raymond}
\affiliation{LIGO - California Institute of Technology, Pasadena, CA 91125, USA}
\author{C. R\"{o}ver}
\affiliation{Albert-Einstein-Institut, Callinstr. 38, 30167 Hannover, Germany}
\author{L. Singer}
\affiliation{LIGO - California Institute of Technology, Pasadena, CA 91125, USA}
\author{M. van der Sluys}
\affiliation{Radboud University Nijmegen, P.O. Box 9010, NL-6500 GL Nijmegen, The
Netherlands}
\author{R.J.E. Smith}
\affiliation{School of Physics and Astronomy, University of Birmingham, Birmingham, B15 2TT, UK}
\author{A. Vecchio}
\affiliation{School of Physics and Astronomy, University of Birmingham, Birmingham, B15 2TT, UK}
\author{J. Veitch}
\affiliation{Nikhef, Science Park 105, Amsterdam 1098XG, Netherlands}
\author{S. Vitale}
\affiliation{Massachusetts Institute of Technology, 185 Albany St, 02139 Cambridge, USA}

\begin{abstract}
The problem of reconstructing the sky position of compact binary coalescences detected via gravitational waves is a central one for future observations with the ground-based network of gravitational-wave laser interferometers, such as Advanced LIGO and Advanced Virgo.  Different techniques for sky localisation have been independently developed. They can be divided in two broad categories: fully coherent Bayesian techniques, which are high-latency and aimed at in-depth studies of all the parameters of a source, including sky position, and ``triangulation-based" techniques, which exploit the data products from the search stage of the analysis to provide an almost real-time approximation of the posterior probability density function of the sky location of a detection candidate. These techniques have previously been applied to data collected during the last science runs of gravitational-wave detectors operating in the so-called initial configuration. 

Here, we develop and analyze methods for assessing the self-consistency of parameter estimation methods and carrying out fair comparisons between different algorithms, addressing issues of efficiency and optimality.  These methods are general, and can be applied to parameter estimation problems other than sky localisation.  We apply these methods to two existing sky localisation techniques representing the two above-mentioned categories, using a set of simulated inspiral-only signals from compact binary systems with total mass $\le 20\,M_\odot$ and non-spinning components.  We compare the relative advantages and costs of the two techniques and show that sky location uncertainties are on average a factor $\approx 20$ smaller for fully coherent techniques than for the specific variant of the ``triangulation-based'' technique used during the last science runs, at the expense of a factor $\approx 1000$ longer processing time. 
\end{abstract}

\maketitle

\section{Introduction}
Ground-based gravitational-wave (GW) laser interferometers -- LIGO \cite{Abbott:2007kv}, Virgo \cite{iVirgo} and GEO-600 \cite{Grote:2008} -- have completed science observations in 2010 (S6/VSR2-3) \cite{ligo12} 
in the so-called initial configuration, and are currently being upgraded with the plan to start running again from 2015 at a significantly improved sensitivity \cite{AdvLIGO, AdvVirgo}. No detection was achieved during this initial period of observations; however, the expectations are that by the time the instruments reach design ``advanced'' sensitivity they shall routinely detect gravitational-wave signals. One of the most promising candidate sources for detection are coalescing binary systems of compact objects containing neutron stars and black holes \cite{LVC2010}. 

One of the key pieces of information to extract is the source location in the sky. Once a detection candidate is identified by search pipelines, the location parameters that describe the source are reconstructed using a number of techniques, both high and low-latency \cite{abadie12,LVC2013a}. In contrast to traditional telescopes, gravitational-wave instruments are all-sky monitors and the source location in the sky is reconstructed a posteriori. Information about the source geometry is primarily encoded in the relative time of arrival of GW radiation at the different detector sites, together with the relative amplitude and phase of the GWs as seen in different detectors. Constraining the source location on the sky will be an important element of the analysis, because it allows for follow-ups of the relevant portion of the sky with electro-magnetic instruments, possibly over a wide spectral range, and could offer information about the environment of a GW-detected binary~\cite{metzger2012,nissanke2011,kelley2010}.  The electro-magnetic signatures associated to the merger of the compact objects are expected to be transient, so the timescale over which sky location information becomes available from the gravitational-wave ground-based network is also important.

For this reason the problem of reconstructing the sky position of GW sources with the network of ground-based laser interferometers is an area of active work in preparation for advanced instruments \cite{Cavalier06,Veitch:2012,Nissanke:2013, kasliwal13, fairhurst2011,grover2013}.  By the end of observations with instruments in initial configuration, 
two main implementations had been used to determine the sky localization uncertainty region of a coalescing binary candidate~\cite{abadie12,LVC2013a}:
\begin{itemize}
\item \texttt{LALInference}~\cite{lalinf_method}, a library of fully coherent Bayesian analysis algorithms, computes the posterior Probability Density Function (PDF) on the sky location and other parameters, on the timescale of hours to several weeks, depending on the specific signal. Using two classes of stochastic sampling techniques, Markov-Chain Monte Carlo \cite{gilks1996,MCMC:2004,vanderSluys2008} and Nested Sampling \cite{Skilling2006,veitch10,Feroz:2008}, \texttt{LALInference} coherently analyses the data from all the interferometers in the network and generates the multi-dimensional PDF on the full set of parameters needed to describe a binary system, before marginalising over all parameters other than the sky location (a binary in circular orbit is described by 9 to 15 parameters, depending on whether spins of the binary components are included in the model).
\item A much faster low-latency technique, that we will call \texttt{Timing++} \cite{abadie12}, uses data products from the search stage of the analysis, and can construct sky maps on (sub-)minute time scales by using primarily time-delay information between different detector sites.   
In particular, the masses, time and phase of arrival, and the amplitude of the signal are searched for in each detector separately and the masses and time of arrival are checked for consistency \cite{Babak:2012zx}. The time of arrival and amplitude of the signal in each detector are the intermediate data products used by \texttt{Timing++} to construct the PDF of the sky location.
\end{itemize}
These two approaches were initially designed to serve different purposes: a thorough parameter reconstruction and a low-latency sky-localisation technique, trading off accuracy for computational speed. 

The goal of this paper is two-fold. Several parameter estimation approaches have been, and continue to be, developed in preparation of the advanced instruments coming on line in 2015. Algorithms may be tuned in specific ways to serve different purposes. The first goal of this paper is to provide fair and rigorous methods to compare different approaches, in order to inform future developments. One of the most actively investigated parameter estimation aspects is sky localisation reconstruction. It is therefore natural to apply these comparison methods to the algorithms used up to now, to check the consistency of the results, quantify relative benefits and identify the areas that need the most attention in the future. The second goal of this paper is to provide the first rigorous comparison of the two sky localisation techniques described above.  
We examine the sky location PDFs for a large number of simulated signals from coalescing compact binaries with total masses up to $20 M_\odot$ in simulated stationary, Gaussian noise.  Although our signal distribution is not astrophysically motivated, it allows us to statistically examine the self-consistency of both techniques by testing whether the claimed uncertainty regions match the actual probability that the source is found at those sky locations.  Furthermore, by comparing the uncertainties in sky location across the code outputs we gain an understanding of the systematic behaviour of each technique.
Many of these comparison methods have now become the routine test-bed in the development effort for gravitational-wave data analysis and may have applicability in other areas of astronomy.

The paper is organised as follows. In Section 2 we describe two techniques used to determine the sky location of a candidate coalescing compact binary. In Section 3, we evaluate the correctness of the two techniques using a simulated population of binaries over a wide range of the parameter space, compare their sky localisation capabilities and latency time scales. Section 4 contains our conclusions and pointers to future work.

\section{Location reconstruction methods}

Gravitational-wave interferometers are, by design, sensitive to sources across much of the sky.  Because of this, position reconstruction estimates rely largely on time delays between sites in a multiple detector network, i.e., triangulation. 
Using only time-delay information, 
there is generally a degeneracy in the position reconstructed. 
For a two-detector network, this degeneracy is a conical surface of constant time delay around the line connecting the two detectors, whose projection onto the sky plane yields a ring.  For a three-detector network this degeneracy is broken into two regions symmetric about the plane defined by the detectors: the intersections of two rings on the sky.  A four (or more) detector network will generally identify a single region in the sky.  However, time-delays are not the only source of sky location information. Though the observed amplitude of gravitational waves depends only weakly on the source location, it typically helps to break these degeneracies in two and three detector networks; further information is contained in the relative phasing between detectors \cite{grover2013}.  In this section we outline the two methods considered so far for position reconstruction.


We can formalise the problem we want to address as follows. The data
\begin{equation}
	\label{eqn:data}
	d_j(t) = n_j(t) + h_j(t; \vec{\theta})\;,
\end{equation}
from each gravitational-wave interferometer in the network $j = 1,\dots, N$, where $N$ is the number of instruments, is a sum of the noise $n_j(t)$ and any signal $h_j(t;\vec{\theta})$, where $\vec{\theta}$ is a vector that describes the set of unknown parameters that characterize the GW source. For this study we consider coalescing binaries of compact objects with approximately circular orbits and negligible spins; $\vec{\theta}$ is a nine-dimensional parameter vector: two mass parameters (the two component masses $m_{1,2}$, or an alternative combination of these, \textit{e.g.}, the symmetric mass ratio $\eta = {m_1m_2}/{\left( m_1+m_2\right)^2}$ and the chirp mass $\mathcal{M} = \eta^{3/5}\left(m_1+m_2\right)$), the distance to the source $D$, the source location in the sky (described by two angles that identify the unit vector $\vec\Omega$ -- \textit{e.g.}, right ascension $\alpha$ and declination $\delta$), the orientation of the binary (polarization $\psi$ and inclination of the orbital plane $\iota$), and the reference phase $\phi_0$ and time $t_0$. To simplify notation, we define 
\begin{equation}
	\label{eqn:param}
	\vec{\theta} = \{\vec\Omega, \vec{\beta}\}\,,
\end{equation}
where $\vec{\beta}$ is the parameter vector that \textit{does not} contain the sky location parameters, right ascension and declination. Regardless of the specific technique that one decides to adopt, the goal is to evaluate $p(\vec\Omega | d)$, the marginalised joint posterior density function of the sky location parameters given the observations. 

A straightforward application of Bayes' theorem allows us to calculate the posterior probability density for a model with parameters $\vec{\theta}$ given the data, $d$, using
\begin{equation}
	\label{eqn:BayesTheorem}
  	p(\vec{\theta}|d) = \frac{p(d|\vec{\theta})\,p(\vec{\theta})}{p(d)} \ .
\end{equation}
The prior probability density, $p(\vec{\theta})$, encapsulates all our {\it a priori} information about the expected distribution of sources in distance, masses or other parameters in the model.  The likelihood $p(d|\vec{\theta})$ is the probability of generating the data set $d$ given an assumed signal with parameters $\vec{\theta}$.  The evidence $p(d)$ is used to normalise the integral of the posterior over the entire parameter space to unity.

\subsection{\texttt{LALInference}}

The evaluation of $p(\vec{\theta}|d)$ is notoriously difficult in high-dimensional problems with complex likelihood functions, as is the case for coalescing compact binaries in a network of laser interferometers.  We have developed a set of sampling algorithms within the LSC Algorithms Library (LAL) \cite{LAL} , collected under \texttt{LALInference} \cite{lalinf_method}, specifically for the analysis of gravitational-wave data, and for what is relevant here, coalescing-binary signal models. The library contains two main stochastic parameter-space exploration techniques: Markov-Chain Monte-Carlo (\texttt{lalinference\_mcmc} \cite{vanderSluys2008}), and nested sampling (\texttt{lalinference\_nest} \cite{veitch10} and \texttt{lalinference\_bambi} \cite{Graff:2012}). Different algorithms are included to validate results during the development stage and to explore a range of schemes to optimise the run time. These techniques have been used to analyse a set of hardware and software injections as well as detection candidates during the last LIGO/Virgo science runs~\cite{LVC2013a}; a technical description of the algorithms will be reported elsewhere~\cite{lalinf_method}.

The output of a \texttt{LALInference} run is a list of ``samples'', values of ${\vec\theta}$ drawn from the the joint posterior probability density function. The density of samples in a region of parameter space is proportional to the value of the PDF. For the specific sky localisation problem we are considering here, the marginalised posterior probability density function on the sky location is simply:
\begin{equation}
	p(\vec\Omega| d) = \int p(\vec\Omega, \vec{\beta} | d) d\vec{\beta}\,,
	\label{eq:lali_post}
\end{equation}
where $p(\vec\Omega, \vec{\beta} | d) \equiv p(\vec{\theta} | d)$ is derived using Eq.~(\ref{eqn:BayesTheorem}). If we could extract an infinite number of samples then we would be able to map out the PDF perfectly; however, these are computationally intensive algorithms, see Section \ref{ss:runTime} for more details, and we typically have $\sim$1000 \textit{independent} samples. The finite number of samples can introduce both stochastic and systematic bias, and so we have implemented a 2-step kD-tree binning process to estimate the PDF that removes the systematic issues \cite{sideryKD}.

The fully coherent Bayesian analysis takes into account the search stage of the analysis only to set the prior range for the arrival time of a gravitational wave around the observed detection candidate. However, the matched-filtering stage of a search already offers processed information that can be used to generate approximate posterior density functions $p(\vec\Omega| d)$. This is the approach taken in \texttt{Timing++}.

\subsection{\texttt{Timing++}}


\texttt{Timing++} ~\cite{abadie12} takes the parameters of the waveform that best fit the data in each detector, as found by the initial search~\cite{Babak:2012zx}, and assumes that the posterior of interest is only a function of the arrival times in each detector, $t^{(i)}$, and the amplitude of the signal in each detector, $A^{(i)}$.  That is, we write
\begin{equation}
p(\vec\Omega | d) \approx p\left(\vec\Omega \middle| t^{(i)},A^{(i)}\right),
\label{firstapprox}
\end{equation}
where $\vec\Omega$ is the location on the sky.  We further assume the information in the arrival times and amplitudes can each be replaced by a single quantity so that
\begin{align}
p(\vec\Omega | d) &\approx p\left(\vec\Omega \middle| t^{(i)},A^{(i)}\right) \nonumber \\
&\propto f(\Delta t_{\rm rss, sc}(\vec\Omega),\Delta A_{\rm rss}(\vec\Omega))\nonumber \\
&\equiv f(\Delta t_{\rm rss, sc},\Delta A_{\rm rss}),\label{secondapprox}
\end{align}
where $f(\Delta t_{\rm rss, sc},\Delta A_{\rm rss})$ is an empirically derived distribution function and $\Delta t_{\rm rss, sc}$ and $\Delta A_{\rm rss}$ are described in the following.  For a source at position $\vec\Omega$, the arrival time at detector $i$ allows us to predict the arrival time at any other fiducial point, which, for the sake of simplicity, we choose to be the geocenter.  In the absence of noise, the predicted geocentric arrival times, computed separately from each detector's measured arrival time, should coincide. The summed squared differences of the predicted arrival times at the geocenter between detector pairs give us a measure of how far we expect to be from the true location:
\begin{equation}
\Delta t_{\rm rss} = \sqrt{\sum_{i> j} \left( \left(t_{\rm ref}^{(i)} - 
t_{\rm geo}^{(i)}(\vec\Omega)\right)- \left(t_{\rm ref}^{(j)} - 
t_{\rm geo}^{(j)}(\vec\Omega)\right) 
\right)^2}, \label{dtrss}
\end{equation}
where $t_{\rm geo}^{(i)}(\vec\Omega)$ is the difference between the arrival time of a signal from  $\vec\Omega$ at detector $(i)$ and at the geocenter, and $t_{\rm ref}^{(i)}$ is the time the signal crosses a {\it reference frequency} in the band of detector $i$. This vanishes in the idealised case of no noise for the true location. By appropriately choosing the reference frequency we minimise the correlation between the determined mass and phase in the waveform, and the recovered time of arrival~\cite{2007CQGra..24S.617A}. This is important since the parameters of the waveform are determined separately in each detector. Moreover, we expect that these errors in timing will scale inversely with the signal-to-noise ratio (SNR) of the system in the high-SNR regime:
\begin{equation}
\Delta t_{\rm rss} = \Delta t_{\rm rss,sc}\frac{10}{\rho},
\end{equation}
where $\rho = \sqrt{\sum_{i}\rho_i^2}$ is the {\it combined SNR}, $\rho_i$ is the SNR measured in detector $i$, and the factor of 10 is chosen as a fiducial SNR.  We use the SNR-corrected $\Delta t_{\rm rss,sc}$ in place of $\Delta t_{\rm rss}$ to remove this dependence on SNR.

Incorporating the amplitude of the signal is more complicated.  The SNR is a function not only of sky location but also of luminosity distance, inclination and polarization of the signal.  Because this method is designed for low-latency sky localisation, a somewhat \textit{ad hoc} measure of amplitude consistency between detectors is used.  The starting point is the fact that 
\begin{equation}
	\rho_i \propto \frac{1}{D_{\rm eff}^{(i)}},
\end{equation}
where $D_{\rm eff}$ is an effective distance, defined by
\begin{equation}
D_{\rm eff} = D\left[F_{+}^2\left(\frac{1+\cos^2\iota}{2}
\right)^2 + F_{\times}^2\cos^2\iota\right]^{-1/2}\,,
\label{deffdef}
\end{equation}
and $F_{+,\times} =  F_{+,\times}(\vec\Omega,\psi)$ are the antenna beam pattern functions, see equations B9 and B10 of Ref.~\cite{2001PhRvD..63d2003A}. While the matched filter detection pipeline produces an estimate of $D_{\rm eff}$ separately in each detector, it is not invertible to obtain any of the quantities in Eq.~(\ref{deffdef}) directly.  With that in mind, we define
\begin{equation}
A^2 \equiv \frac{1}{F_{+}^2(\vec\Omega,\psi=0) +
F_{\times}^2(\vec\Omega,\psi=0)},
\end{equation} 
and use
\begin{equation}
\Delta A_{\rm rss} = \sqrt{\sum_{i > j} \left(\frac{D_{{\rm eff}}^{(i)\,2} - 
D_{{\rm eff}}^{(j)\,2}}{D_{{\rm eff}}^{(i)\,2} + D_{{\rm eff}}^{(j)\,2}} -
\frac{A^{(i)\,2} - A^{(j)\,2}}{A^{(i)\,2} + A^{(j)\,2}}\right)^2},\label{dArss}
\end{equation}
as a measure of the consistency of the calculated and observed difference in response functions between each detector pair.
In contrast to Eq.~(\ref{dtrss}), this quantity is typically not zero in the absence of noise as $A^2 = D_{\rm eff}/D$ only when inclination and polarisation are both 0.  However, the use of amplitude reconstruction in this manner has been determined empirically to improve position reconstruction estimates. In contrast to $\Delta t_{\rm rss,sc}$, there is no adjustment for SNR in $\Delta A_{\rm rss}$. 
\citet{grover2013} showed that phase consistency between detectors can provide additional information on sky location and significantly reduce sky localisation uncertainty; however, phase consistency was not included in \texttt{Timing++}.

Putting together our previous assumptions 
\begin{align}
p(\vec\Omega | d) &\approx p\left(\vec\Omega \middle| t^{(i)},A^{(i)}\right) \nonumber\\
&\propto p(\vec\Omega)f(\Delta t_{\rm rss, sc},\Delta A_{\rm rss}) \nonumber\\
&\approx p(\vec\Omega) f_t(\Delta t_{\rm rss, sc}) f_A(\Delta A_{\rm rss}),\label{fullpostapprox}	
\end{align}
where $p(\vec\Omega)$ is the prior on sky location and in the third line we have assumed that $f(\Delta t_{\rm rss, sc},\Delta A_{\rm rss})$ can be written as the product of two other empirical distributions, $f_t(\Delta t_{\rm rss, sc})$ and $f_A(\Delta A_{\rm rss})$.  In this work we assume isotropic priors on sky location.  In the low-latency search for compact binaries and associated electromagnetic counterparts for which \texttt{Timing++} was designed, a restrictive prior that limited consideration to only areas of the sky containing galaxies was imposed, as described in~\cite{abadie12}.  In practice,  $f_t(\Delta t_{\rm rss, sc})$ and $f_A(\Delta A_{\rm rss})$ are measured beforehand using simulations, where $\Delta t_{\rm rss,sc}$ and $\Delta A_{\rm rss}$ are computed from the recovered arrival times and effective distances, respectively, and the true (known) sky location, $\vec\Omega_{\rm true}$.  This amounts to evaluating $\Delta t_{\rm rss,sc}$ ($\Delta A_{\rm rss}$) according to Eq.~(\ref{dtrss}) (Eq.~(\ref{dArss}))  at $\vec\Omega_{\rm true}$ using the time of arrival (effective distance) from the matched filter pipeline.  A kernel density estimator is then used to estimate the distribution of these quantities. When a candidate is found, $\Delta t_{\rm rss,sc}$ and $\Delta A_{\rm rss}$ are computed across a fixed grid on the sky, and the likelihood is taken from the previously simulated distributions and the result is normalized, leading to an inherently fast method.

\section{Testing \label{testing}}

The goal of this study is to compare the relative performances in terms of sky localisation of \texttt{Timing++} and \texttt{LALInference} and in doing so to develop a set of criteria and general tools that can be applied to many parameter estimation problems in which different techniques are considered. The tests should ensure that each algorithm separately is self-consistent, and then provide fair methods of making comparisons.

For the specific problem considered in this paper, \texttt{Timing++} and \texttt{LALInference} both evaluate the posterior probability density function $p(\vec\Omega | d)$, see Eqs.~(\ref{eq:lali_post}) and~(\ref{fullpostapprox}). For a given model assumption and data realisation, there is an exact PDF of which the algorithms produce an approximation. 
There are many effects that can distort the recovered PDF from the true one. They can be grouped in two broad categories. 

Irrespective of the algorithm that is used, the assumptions on the elements that enter the PDF calculation may differ from the actual problem, and therefore produce a bias in the results. For the problem at hand, they can be summarised as follows: (i) the model waveform family does not describe the actual signal contained in the data; (ii) the noise model is incorrect, and (iii) the choice of priors does not match the actual ones that describe the problem, and in the specific case considered here, the priors from which the source parameters have been drawn. Each of these enter the calculation of the PDF, see Eq.~(\ref{eqn:BayesTheorem}). In the test described here, 
the signal model (the waveform family) is exactly known, and the same waveform family is used for the signal generation and the likelihood calculation. The statistical properties of the noise -- Gaussian and stationary drawn from a known distribution -- are also known. It is however important to emphasise that in the case of \texttt{LALInference} the noise power spectral density (PSD) is estimated from the data surrounding the signal, and as a consequence it does not exactly describe the distribution from which noise is drawn. 
For the \texttt{Timing++} analysis, on the other hand, the noise PSD is taken to be exactly the one used to generate the noise realisations.

A different set of effects that can affect the recovered PDF are more fundamentally intrinsic to the algorithms: (i) the assumptions that go into the likelihood calculation are not perfect,
(ii) there are algorithmic issues that produce errors, and (iii) PDFs cannot be reconstructed perfectly from a finite number of samples (post-processing). 
The likelihood calculation makes assumptions about the form of the noise and so is linked to the previously mentioned noise issue. 
For \texttt{Timing++}, the likelihood is calculated using a mix of approximations and simulated runs. This is a point of possible bias entering the results of the \texttt{Timing++} runs; measuring its extent is part of our investigation.

As well as the obvious statement that the algorithm must be working correctly, it was found with \texttt{LALInference} that the way that the results are processed to create a continuous PDF from discrete samples from the posterior can also introduce noticeable distortions. This is linked to the finite sampling issues mentioned previously and fixed with 2-stage kD-trees \cite{sideryKD}.

While in theory the sources of bias due to the test itself are straightforward to control, any erroneous results may either be due to code issues or a failure to properly treat the setup issues; both of which may give very similar distortions in the final PDF. This leads to a cycle of code checking and test setup checking while codes are being developed.
This is particularly true of the \texttt{LALInference} type algorithms that, with the correct setup, should precisely recover the PDF, creating a stringent checking mechanism for the codes' self-consistency.

\subsection{Test population}\label{sec:population}

To set up a rigorous comparison test-bed we have 
considered 360 mock inspiralling compact binary signals from a population of binary sources and ``injected'' the waveforms
into Gaussian and stationary noise representing observations with the 3-detector network consisting of the two LIGO
detectors at Hanford, WA and Livingston, LA and the Virgo detector
near Pisa, Italy. The power spectrum of the noise was chosen to
mimic the LIGO sensitivity achieved during the last science run~\cite{ligo12}, and was the same for all the instruments of the network, including Virgo. A subset of this population has been recently used for other parameter estimation studies, see Refs.~\cite{grover2013,rodriguez13}. The noise data were generated with the infrastructure used for the NINJA-2 project~\cite{ninja2-mdc}. The low-frequency cut-off was set to 40 Hz.

The source distribution was chosen to test these two sky localisation approaches over a large range of signal-to-noise ratios and physical parameters that describe stellar mass binary systems, rather than being astrophysically motivated. The mass distribution was uniform in component masses with $1\,M_\odot \le m_{1,2} \le 15\,M_\odot$ and a cut off on the total mass $m_1 + m_2 \le 20\,M_\odot$. The sky position and orientation of the systems with respect to the interferometers were distributed uniformly. For distances between 10 and 40 Mpc the logarithm of the distance was uniformly distributed in order to give a broad range of network SNRs above the detection threshold.

The waveforms used to generate, and then analyse, the signal are restricted post-Newtonian approximations of the inspiral phase, with spins of the binary components set to zero. The time-domain TaylorT3 and TaylorT4 approximants of the LSC Algorithm Library (LAL)~\cite{LAL} at second post-Newtonian order in phase, in which the differential equations that describe the evolution of a characteristic orbital velocity and phase of the system are Taylor expanded in terms of the characteristic velocity of the two inspiralling objects~\cite{buonanno09}, were used for \texttt{Timing++} and \texttt{LALInference}, respectively. The precise forms of the two families of waveforms have phase differences from  post$^{2.5}$-Newtonian order and above, which has no effect for the purpose of these comparisons; the crucial factor for these tests was that each code used the same waveform family for injection and subsequent recovery of the signal. It was necessary to use different waveforms in each code due to compatibility issues of the implementations.

\begin{figure}
  \includegraphics[width = 8cm]{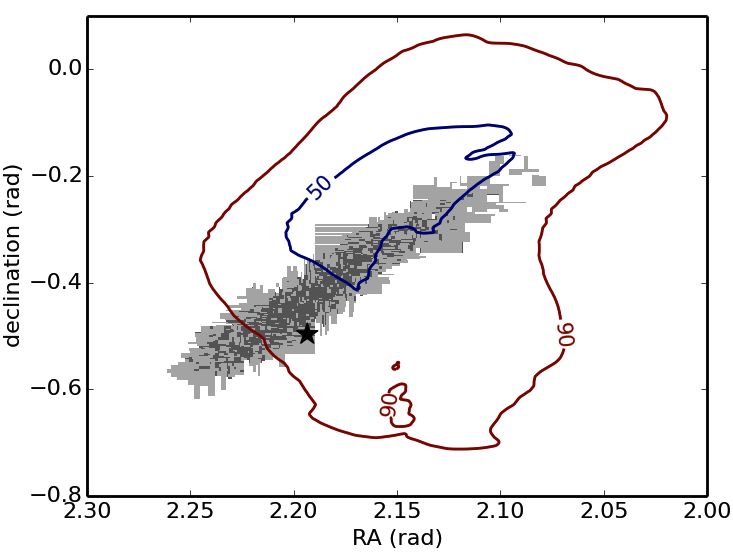}
  \caption{\label{fig:skyMap} An output PDF of the sky position from the two codes. The contour lines label the 50\% and 90\% credible regions for \texttt{Timing++} while the light and dark shaded regions show the 50\% and 90\% credible regions respectively for \texttt{LALInference}. The star indicates the source location.}
\end{figure}
The synthetic data containing GW signals added to noise were processed using the standard matched-filter search pipeline \texttt{ihope}~\cite{Babak:2012zx} used in the LIGO/Virgo analyses in this parameter range, see, \textit{e.g.}, Ref.~\cite{PhysRevD.85.082002} and references therein. \texttt{LALInference} was run on all the 360 injections, with a flat prior on the time of arrival over a range of $\pm 100$ ms around the time of the injection.  
\texttt{Timing++} uses an additional criterion that the SNR must be greater than 5.5 in each of three detectors; 243 candidates passed this cut. Figure \ref{fig:skyMap} gives an example output PDF from one of the runs. For the self-consistency tests described in Section~\ref{ss:self_consistency} we used all the results available for each algorithm. For the comparisons between the codes in Section~\ref{ss:comparisons}, we only used those data sets for which results from both methods are available.

\subsection{Self-consistency checks}
\label{ss:self_consistency}

We describe the PDF via credible levels (CL): the integrated probability, in our case $p(\vec\Omega|d)$, over a given region of the parameter space. In particular we consider the smallest region, or minimum credible region (CR$_\mathrm{min}$), for a given CL; in our case, this corresponds to the smallest region in the sky that contains the given probability that the source is in that location. More formally, for a given CL, any credible region (CR) must satisfy
\begin{equation}
  \mathrm{CL} = \int_{\mathrm{CR}} p(\vec\Omega|d) d \vec\Omega\,.
\end{equation}
We can then find the smallest region such that this still holds, which we call CR$_\mathrm{min}$.
By considering the full range of probabilities we can map out the PDF with a set of contours that bound each CR$_\mathrm{min}$.

While the analysis of a single GW signal will not tell us very much about the correctness of the analysis, considering how CL and CR$_\mathrm{min}$ are related over a large number of GW signals gives us statistical information: Does a given credible level really correspond to the probability of finding the source in that location? For each run and a given CL we can check if the injection's parameter coordinates fall within the associated CR$_\mathrm{min}$;
if there are no sources of bias in the analysis, this should happen with probability CL in order for the credible regions to be meaningful.  
We can plot a cumulative figure, over all injected signals and the full range of CLs, of the proportion of injections found within a given CL's CR$_\mathrm{min}$.  We expect this to be diagonal, up to statistical fluctuations arising from a finite number of injections.  Deviations from the diagonal indicate that the parameter-estimation algorithm does not correctly evaluate the PDF, or other sources of bias are present,  {\it e.g.}, the priors used in the analysis do not match the distribution of the injected source population.

\begin{figure}
  \includegraphics[width = 8cm]{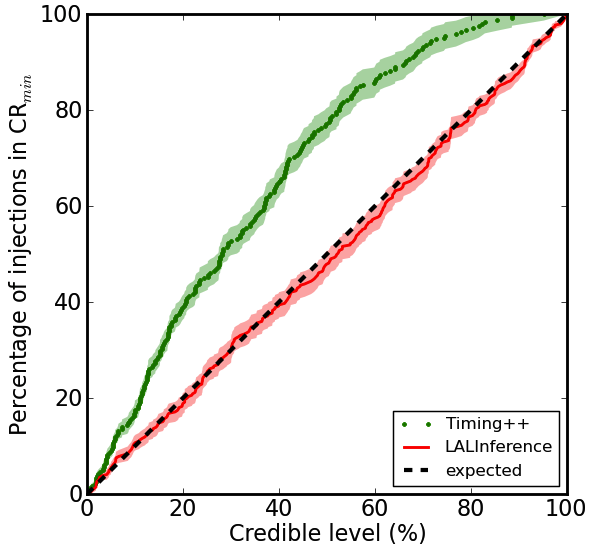}
  \caption{\label{fig:trueVsCalculatedCL} For each credible level (CL) we plot the number of injections that fall within the associated minimum credible region CR$_\mathrm{min}$ for all the signals analysed with \texttt{LALInference}, bottom (red) curve, and \texttt{Timing++}, top (green) curve. The error bars correspond to the binomial error, see text for more details. A self-consistent algorithm gives results that lie along the diagonal line of this plot. Results that fall above the expected line, as is the case for \texttt{Timing++}, highlight an algorithm that is overcautious in its estimation of CR$_\mathrm{min}$.}
\end{figure}

The results of this test from all the signals detected out of the 360 injections in each of \texttt{LALInference} and \texttt{Timing++} is shown in Figure \ref{fig:trueVsCalculatedCL}.
The error bars are calculated from the expected variance in the number of injections that fall within a given CR.
For a CL of $p$, and $n$ runs, the variance on the number of sources found within CR$_\mathrm{min}$ is $np(1-p)$ if the fraction of injections that fall within a given CR$_\mathrm{min}$ is really described by the binomial distribution, as expected. The error bars on the fraction of injections found within a given CR$_\mathrm{min}$ are given by the standard deviation normalised by the number of runs, $\sqrt{p(1-p)/n}$.

We can see here that \texttt{LALInference} produces results that indeed follow the expected relation; we can therefore conclude that the algorithm is self-consistent. During the \texttt{LALInference} development, parallel to this investigation, this test has been used as one of the primary tools to check the algorithm. As well as checking sky location, this test was done in each of the model parameters separately, though rather than using the minimal CL it is easier and sufficient to use a connected credible region whose lower bound is the lowest value of the parameter being investigated.

On the other hand, the results obtained with \texttt{Timing++} show a significant deviation from the expected behaviour: the calculated CRs for \texttt{Timing++} do not represent the `true' CL. As the results are \textit{above} the expected behaviour, the sky regions are too large. This shows that \texttt{Timing++} is not `self-consistent'. This is not necessarily unexpected, because \texttt{Timing++} is purposefully an approximation in favour of speed; it is useful to note that \texttt{Timing++} is over-conservative. 

From these results it also follows that we need to be cautious when designing comparisons between \texttt{Timing++} and \texttt{LALInference} applied to the same GW signal. We consider these comparisons in the next section.

\subsection{Comparisons}
\label{ss:comparisons}

We can now turn to comparisons between \texttt{Timing++} and \texttt{LALInference}, and we consider two different figures of merit for this.

For a self-consistent code, the CR$_\mathrm{min}$ of a chosen CL is a natural metric of the ability of the algorithm to localise the source. This is equivalent to stating the expected smallest region of the sky that needs to be scanned by a follow-up observation to have a given probability that the actual source location is covered. Here we will consider the 50\% minimum credible region, and therefore set CL $= 0.5$. While this is natural for the fully coherent Bayesian codes, the same is not true of \texttt{Timing++}. We saw in the previous section that \texttt{Timing++} is not self-consistent: it does not provide the `correct' CRs at a given CL, but actually overstates it. 

It is however still interesting and possible to know the size of the CR$_\mathrm{min}$ that relates to the `true' CL.  From the self-consistency test we have a relation between the output CRs and the `true' CLs from \texttt{Timing++}. This means we can compare the output areas of the minimal credible regions of the `true' 50\% CL by using the quoted 23\% CR$_\mathrm{min}$ from \texttt{Timing++} and the 50\% from \texttt{LALInference}. In other words, we are correcting for the lack of self-consistency of \texttt{Timing++} and can produce a fair comparison of the two methods.

\begin{figure}
  \includegraphics[width = 8cm]{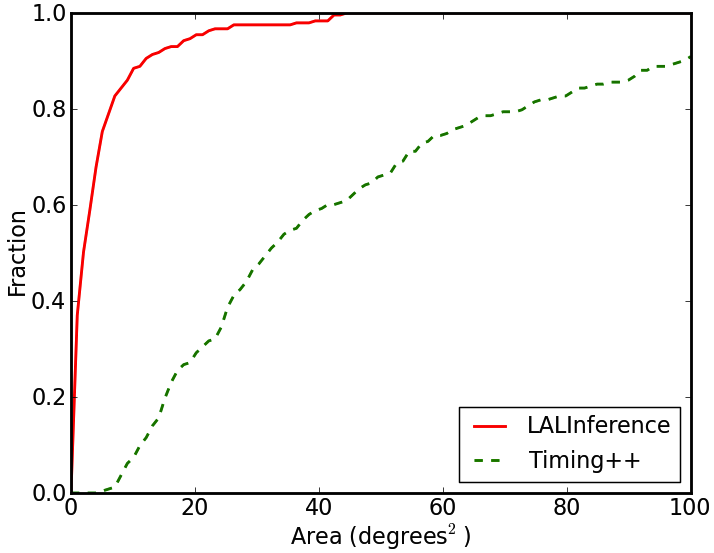}
  \caption{\label{fig:cumuAreaFixed} The fraction of detected signals whose associated true (corrected) 50\% CR$_\mathrm{min}$ covers less than a given area on the sky. We can see that \texttt{LALInference} gives much tighter constraints than \texttt{Timing++} on the location of a source.}
\end{figure}

Figure~\ref{fig:cumuAreaFixed} shows the fraction of signals whose 50\% CR$_\mathrm{min}$s were smaller than a given area.
We can see that even after the corrections to the CLs are implemented, \texttt{Timing++} gives significantly larger CR$_\mathrm{min}$s. This happens because the PDFs returned from \texttt{Timing++} are not quite the same shape as the `correct' PDFs that \texttt{LALInference} is returning; the differences are not simply a rescaling of the width of the peak.

While this test was quite natural from the Bayesian framework point of view, another piece of information that would be passed to followup telescopes would be a list of the most likely ``pixels'' on the sky. One can easily consider a follow-up strategy in which these tiles are observed in order with telescopes until a possible counterpart of the GW-detected source is imaged (or one runs out of pointings). This searched area is equivalent to the size of the CR$_\mathrm{min}$ whose boundary passes through the source's true location on the sky. Furthermore, by considering this area for both approaches we bypass the need to correct for the true relation between probability and CL.
\begin{figure}
  \includegraphics[width = 8cm]{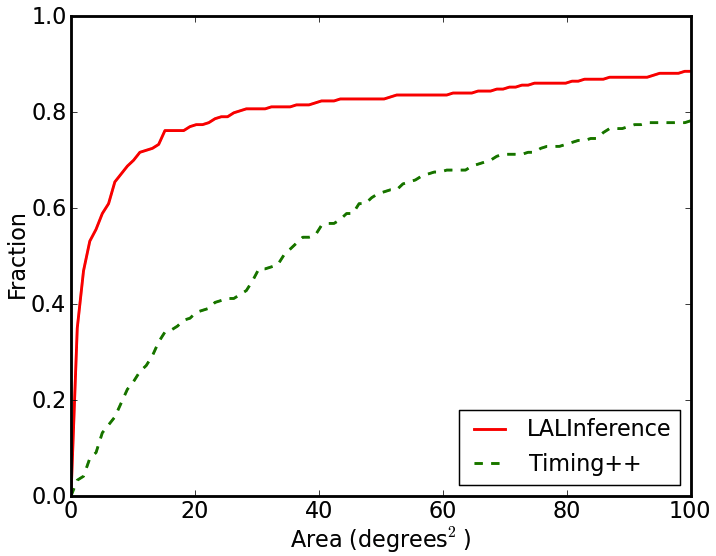}
  \caption{\label{fig:cumulativeInjArea} The fraction of sources where the injection would have been imaged after searching less than the given area in a telescope greedy algorithm.
  }
\end{figure}
Figure~\ref{fig:cumulativeInjArea} shows the fraction of sources that would be imaged after only the given area is searched over, for each source, using the CR$_\mathrm{min}$s as discussed above. We can see that there is a significant difference between the two sky localisation approaches; for example, 76\% of sources would be found after
searching 20 deg$^2$ if we followed the output of \texttt{LALInference}, where as we would only have found 38\% of the injections by
following \texttt{Timing++}.

\begin{figure}
  \includegraphics[width = 8cm]{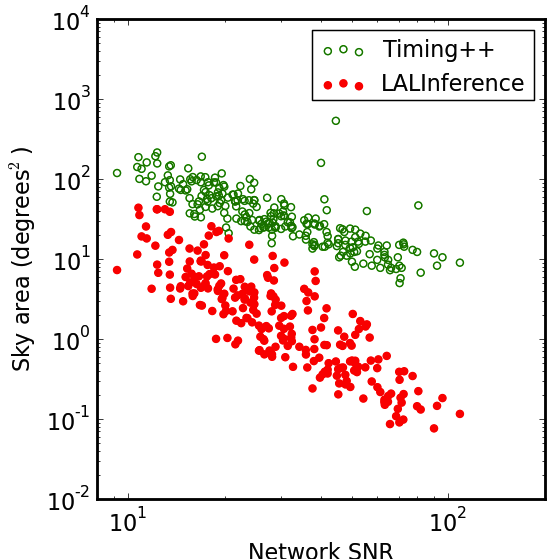}
  \caption{\label{fig:scatterCLArea} The sky area of the 50\% true (corrected) minimum credible region for each of the sources as a function of the optimal network SNR of the signal.  While there is some scatter, the areas from \texttt{LALInference} (solid (red) dots) scale as $\propto 1/$SNR$^2$, as one would expect, while the areas from \texttt{Timing++} (open (green) circles) are closer to $\propto 1/$SNR.}
\end{figure}

\begin{figure}
  \includegraphics[width = 8cm]{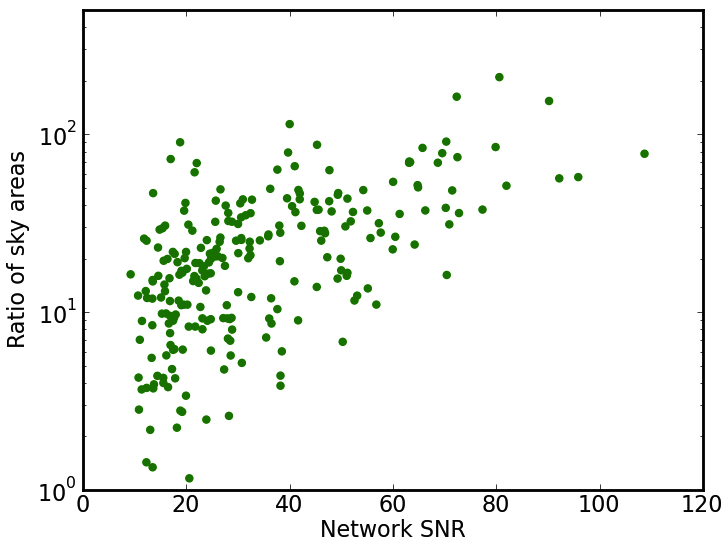}
  \caption{\label{fig:scatterCLAreaFraction} The ratio
    of recovered areas of the 50\% true (corrected) CRs using \texttt{LALInference} as the
    baseline. While there is some scatter, \texttt{LALInference} is consistently producing smaller areas than \texttt{Timing++} by a factor which is roughly 10 for low SNRs and approximately scales with SNR.}
\end{figure}

To gain a better feel for the difference in the calculated areas for the two methods we compared the areas injection-by-injection. We plot the areas of the true (corrected) 50\% CR found by each code where the
injections are sorted by SNR (Figure \ref{fig:scatterCLArea}). For the \texttt{LALInference} results we can
see the expected scaling of the $\textrm{area} \propto  1/\textrm{SNR}^2$.
We also plot the ratio of the 50\% CR$_\mathrm{min}$ areas determined by the two codes in Figure \ref{fig:scatterCLAreaFraction}.
We can see that there is significant spread around the typical factor of
20 difference between the calculated CR$_\mathrm{min}$ areas.

These results should not be taken as a statement on the expected sky localisation accuracy as the underlying injection distribution
is not astrophysical. The set of injections was chosen to test and compare
the codes over a wide region of parameter space and should be treated as such.

\subsection{Run time}
\label{ss:runTime}

\texttt{Timing++} has been set up with speed in mind and so the run time to extract the sky location after data is received is on the order of minutes \cite{abadie12}. Prior to the analysis, the distributions $p(\Delta t_{\mathrm{rss,sc}} | \vec\Omega)$ and  $p(\Delta A_\mathrm{rss} | \vec\Omega)$ need to be generated, and this is done with large scale simulations. Despite being computationally expensive -- the simulations require on the order of days to weeks -- this step is done prior to the actual analysis and therefore has no impact on the latency of the on-line analysis.


While considering code speed, we need to specify the specific sampler used in \texttt{LALInference}. Here we report results for \texttt{lalinference\_mcmc}, the sampling method that was used for this study. A comparison between different samplers in \texttt{LALInference} will be reported elsewhere.

There are two main metrics of computational cost that we consider here: the so-called ``wall time'' (the time an analysis job takes from start to finish), and the total processing (CPU) time. \texttt{lalinference\_mcmc} is designed to take advantage of multiple cores and runs in parallel on different processors.  The parallel chains explore likelihoods at different contrast levels (``temperatures'').  We find that roughly $10$ chains are optimal for improving sampling and convergence for the data sets considered in this study; therefore, CPU times are a factor of ten larger than wall times.


The important quantity to report for \texttt{LALInference} is the time required to output a new independent sample of the posterior PDF. The precise number of samples that we deem necessary to describe the PDF is a balance between speed and precision; as mentioned earlier, finite sample-size issues are a concern for post-processing, and we have found that we require at least 1000 independent samples. 

\begin{figure}
  \includegraphics[width = 8cm]{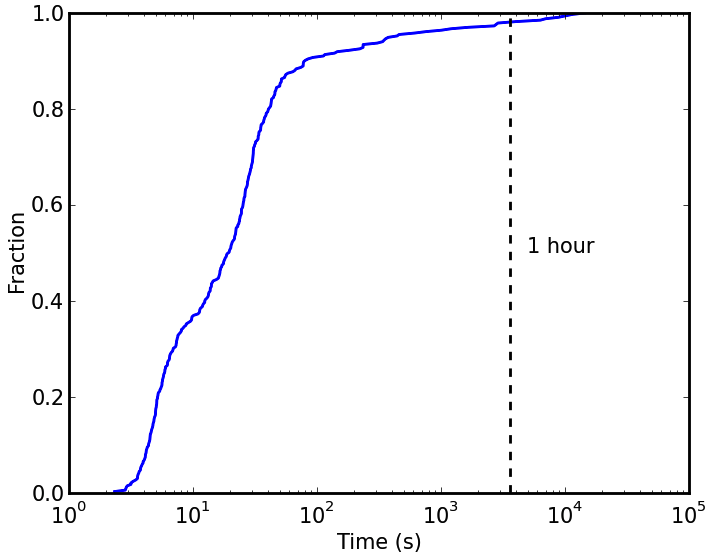}
  \caption{\label{fig:MCMCspeed} The cumulative distribution of wall times for \texttt{lalinference\_mcmc} to output a new independent sample across the runs performed to generate the results reported in this paper.  With 10 cores used for each run, CPU times were a factor of 10 larger.}
\end{figure}

In Fig.~\ref{fig:MCMCspeed} we show the fraction of the analysis runs that output a single independent sample within a given wall time. This quantity was derived by dividing the total wall time of each injection run by the number of independent samples generated in that run. 
From this graph we can see that 90\% of the runs had output 1000 independent samples in $\sim 14$ hours of wall time. The runs were done on nodes composed
of Intel Nehalem E5520 processors (2.26 GHz) with Infiniband DDR interconnects. 

\section{Discussion}

In this paper we have considered two sky localisation algorithms, \texttt{LALInference} and \texttt{Timing++}, used during the final science run of the LIGO and Virgo instruments in initial configuration. Our goal was to assess the relative benefits and costs of the two approaches, and to develop a strategy as well as practical tools to evaluate the consistency of the results and inform the future direction of development. We are now applying these tools to a number of parameter-estimation research projects. 

For the study presented in this paper we have considered a synthetic data set representing a three-detector network.  GW signals generated during the inspiral phase of the coalescence of binary systems with total mass smaller than $20\,M_\odot$ and non-spinning components were added to Gaussian and stationary noise representative of the sensitivity of initial LIGO. We have chosen the range of source parameters in order to best explore the performance of the algorithms. This is important for testing purposes, but one cannot draw conclusions about the actual performance of the GW instruments in future observations from these simulations. To address that question, one would need to consider an astrophysically motivated population of sources, {\it e.g.}, binaries distributed uniformly in volume, and then consider sky localisation only for those signals that pass a detection threshold of the search pipeline. 

As discussed in Section \ref{testing}, posteriors can be systematically biased because of incorrect models, inaccurate priors, insufficient sampling, or improper post-processing to estimate credible regions.

Incorrect models are always a concern in parameter estimation.  Our likelihood model, $p(d|\vec\theta,H)$, could be incorrect because of inaccuracies in the waveform models, noise models, or calibration errors.  Waveforms may not include certain features, (\textit{e.g.}, in this study, we did not allow for spinning binary components) or are affected by limitations in the accuracy of waveform models; efforts are under way to develop more accurate and complete models \cite{Hannam:2013waveform,Taracchini:2013} and to account for waveform uncertainty directly in parameter estimation. 
Real detector noise is neither stationary nor Gaussian; promising strides have been made in accounting for noise non-stationarity \cite{Littenberg:2013}, shifts in spectral lines, and even glitches in the noise.  The impact of calibration errors on parameter estimation was analyzed in the context of initial detectors~\cite{Vitale:2012}; this analysis will need to be repeated for advanced networks.  In this study, our models were correct by construction, as we used stationary, Gaussian noise, assumed perfect calibrations, and employed the same waveform families for injections and templates.

In this paper, we explicitly made sure that the priors assumed by \texttt{LALInference} were identical to the injection distribution to guarantee that inaccurate priors did not introduce a bias in the results, and our code development efforts and thorough testing ensured that insufficient sampling was not a concern.  

We did find early in our studies that our initial approach to post-processing could lead to systematically understated posterior credible regions.  We addressed this by developing a more sophisticated post-processing procedure (see below and \cite{sideryKD}).

There is an important difference between self-consistency and optimality of the results. Self-consistency is a requirement of any code that claims to provide reliable credible regions: the credible regions corresponding to a given confidence level must include the true source parameters for a fraction of signals equal to that confidence level.  Optimality refers to an algorithm's ability to return the smallest credible region among all self-consistent credible regions.  A self-consistent algorithm need not be optimal.
When it comes to our ability to optimise, we must consider both the main algorithm, and the post-processing of the results.

As has been shown here, the proportion of available information that is utilised in the analysis can significantly affect the accuracy of parameter estimation. \texttt{LALInference} uses the data taken from all detectors coherently and thereby recovers small credible regions while staying self-consistent. \texttt{Timing++} on the other hand purposefully makes simplifications, using intermediate data products from the incoherent analysis of individual detector data, and hence the recovered credible regions, even after a correction for self-consistency, are much larger. The trade-off lies in the runtime of the analyses: \texttt{Timing++} returns a sky location within minutes from the completion of the search, whereas \texttt{LALInference} takes approximately half a day (wall time) for the specific waveform family and network considered here. 

Optimality is also important for the post-processing of the algorithms' output to generate marginalised PDFs and credible regions. A binning scheme is traditionally applied, in which the parameter space is split into a uniform grid and the average density of samples in each region found. Using a greedy approach based on this scheme to calculate optimal credible regions (CR$_\mathrm{min}$), self-consistency is broken \cite{sideryKD}. For \texttt{LALInference} we have therefore implemented a more sophisticated way of setting up the initial bins known as a kD-tree so that the resolution of bins follows the density of the samples. A two-stage approach to ordering bins and estimating their contribution to the posterior is required to satisfy self-consistency while managing to get close to optimality. This method will be described in full elsewhere \cite{sideryKD}.
While we have successfully applied this to 2-dimensional posteriors in this study, we cannot currently extend this scheme to higher dimensions: the number of \texttt{LALInference} output samples required for accurate kD-tree PDF interpolation grows exponentially with the number of dimensions and so the runs become impractically long.

While we have outlined the procedure for testing that an algorithm and its implementation report self-consistent results, it is difficult to check for optimality. One approach is to set up runs where the posterior PDFs are known, which was indeed done as part of the \texttt{LALInference} testing and validation \cite{lalinf_method}. By design these will be simple analytic functions and there is no general prescription that will test for all circumstances.

The work that we have reported here, and the tools that we have developed and described have already been important in the further development of \texttt{LALInference}. A new low-latency sky local localisation has also been developed~\cite{bayes_star}.
It is important for future work that while we strive to improve on our methods in both speed and accuracy, we continue to validate these methods against the tests described here in order to have a reliable analysis when the next generation of detectors begins collecting data.

\section*{Acknowledgements}
JV was supported by the research programme of the Foundation for Fundamental Research on Matter (FOM), which
is partially supported by the Netherlands Organisation for Scientific Research (NWO).
NC was supported by the NSF grant PHY-1204371.
PG was supported by a NASA Postdoctoral Fellowship from the Oak Ridge Associated Universities.
BF, WF, and VK were supported by NSF grant PHY-1307020, and BF was also
supported by NSF grant DGE-0824162.
VR was supported by a prize postdoctoral fellowship from the California Institute of Technology division of Physics, 
Mathematics \& Astronomy and LIGO Laboratory.
ROS was supported by NSF awards PHY-0970074 and the UWM Research Growth Initiative.

\bibliography{../../../bibtex/cbc-group,skyloc}

\end{document}